\documentclass[10pt,letterpaper]{article}
\usepackage{OSAmeet, palatino, geometry, textcomp, amsmath, graphicx, color}
\usepackage{microtype}
\usepackage{color}

\newcommand{\eh}[1]{\medspace\textrm{#1}}

\begin{document}
\begin{center}
	\textbf{\LARGE{Observation of supersymmetric scattering\\
	               in photonic lattices}\\}
			\vspace{5mm}
	\textbf{Matthias Heinrich$^{1,2,\ast}$
			Simon St\"utzer$^{2}$,
			Mohammad-Ali Miri$^{1}$,\\
			Stefan Nolte$^{2}$,
			Demetrios N. Christodoulides$^{1}$ and
			Alexander Szameit$^{2}$}\\
			\vspace{5mm}
	\small{\textit{$^1$\,CREOL, The College of Optics and Photonics,
						 University of Central Florida,\\
						 4304 Scorpius St., Orlando FL-32816, USA\\
			\vspace{1mm}
				   $^2$\,Institute of Applied Physics, Abbe Center of Photonics,
			       		 Friedrich-Schiller-Universit\"at Jena,\\
			       		 Max-Wien-Platz 1, D-07743 Jena, Germany\\
			\vspace{1mm} 
			 	   $^\ast$\,e-mail: {\color{blue}matthias.heinrich@uni-jena.de}}}
\end{center}

{\noindent \hskip.35in\begin{minipage}{5.75in} \parindent.5in% 
    \noindent \normalsize \textbf{Abstract:}
    Supersymmetric (SUSY) optical structures display a number of intriguing
	properties that can lead to a variety of potential applications, ranging from
	perfect global phase matching to highly efficient mode conversion and novel
	multiplexing schemes. Here, we experimentally investigate the scattering
	characteristics of supersymmetric photonic lattices. We directly observe the
	light dynamics in such systems and compare the reflection/transmission
	properties of SUSY partner structures. In doing so, we demonstrate that
	discrete settings constitute a promising testbed for studying the different
	facets of optical supersymmetry.\vspace{2mm}\\
	{\small\textbf{OCIS codes:} \textit{290.0290}~(Scattering);   
    							\textit{230.7370}~(Waveguides);
    							\textit{260.2710}~(Inhomogeneous optical media)}
\end{minipage}}

\noindent
The concept of supersymmetry was originally developed in quantum field theory as
a means to unify the mathematical treatment of bosons and fermions
\cite{RamondSUSY}. While to this date no evidence for SUSY behavior has been
found in nature, the underlying theoretical formalism can be adapted in a
variety of different fields of physics, ranging from nonrelativistic quantum
mechanics \cite{CooperKhare} to the design of semiconductor heterostructures
\cite{SUSYQuantumWell} and the optimization of frequency conversion in quantum
cascade lasers, to name a few \cite{BaiCitrin}. Recently, optical SUSY was found
to provide a systematic approach to phase matching and mode manipulation in
wave-guiding structures \cite{SUSY-PRL,SUSY-PRA,UkraineSUSY}. Along these lines,
supersymmetric mode converters have been proposed as versatile building blocks
for readily scalable, highly efficient and fully integrated mode division
multiplexing arrangements \cite{SUSY-MC}.

Yet, the consequences of supersymmetry are by no means limited to guided waves.
The very principles, which serve to establish perfect global phase matching between
waveguides linked by supersymmetric transformations, can also have a profound
impact on the scattering characteristics. As it turns out, two SUSY partner
structures exhibit the same intensity reflection and transmission coefficients
for all angles of incidence. This peculiar behavior could render two apparently
dissimilar structures virtually indistinguishable, suppress reflections at their
interface, or make them altogether invisible to a remote observer
\cite{SUSY-PRA, LonghiReflection,
LonghiInvisible1,LonghiInvisible2,MidyaInvisible}. Based on these
considerations, it is even possible to supersymmetrically extend transformation
optics to enable the design of low-index dielectric equivalent structures that
faithfully mimic the optical properties of high-contrast, or negative-index,
arrangements \cite{SUSY-TO}.

In this work, we experimentally investigate and report the first observation of
light scattering from SUSY structures. Supersymmetric arrangements are realized
in coupled networks of photonic lattices fabricated via femtosecond laser
writing schemes \cite{WaveguideReview}. The corresponding light dynamics in such
systems are directly visualized by means of waveguide fluorescence microscopy
\cite{QuasiIncoherent}. Our findings extend the scope of SUSY experiments beyond
the previously investigated bound state scenarios \cite{SUSY-MC}, and in doing
so, illustrate how discrete arrangements readily allow for the implementation and observation of scattering-invariant transformations mediated by optical supersymmetry.

In optics, supersymmetric transformations involve a factorization of the
Hamiltonian-like operator $\mathcal{H}=-\partial_x^2 -k_0^2n^2(x)$, which
governs the propagation of light in a continuous refractive index landscape
$n(x)$ in the $x-z$ plane \cite{SUSY-PRL}. Here, $k_0$ represents the vacuum
wave number. For a Hermitian conjugate pair of operators $\mathcal{A}$ and
$\mathcal{A}^\dagger$, one can readily see that the Hamiltonian
$\mathcal{H}^{(1)}=\mathcal{A}^\dagger\mathcal{A}$ shares a common set of
propagation constants, or eigenvalues $\lambda$, with
$\mathcal{H}^{(2)}=\mathcal{A}\mathcal{A}^\dagger$. Moreover, is known that for
any incident angle $\theta$, the complex reflection and transmission
coefficients $r^{(1,2)}$ and $t^{(1,2)}$ associated with the corresponding index
landscapes $n^{(1,2)}$ are equal up to a phase \cite{SUSY-PRL}. Intensity-wise,
the respective reflectivities $R^{(1)}=R^{(2)}=|r^{(1,2)}|^2$ and
transmittivities $T^{(1)}=T^{(2)}=|t^{(1,2)}|^2$ of these two structures are
strictly identical.

The conceptual framework of SUSY can be naturally extended to discrete systems
such as periodic arrays of evanescently coupled waveguides, or ``photonic
lattices''. In the coupled-mode approximation, the evolution of wave packets in
this latter type of arrangements is described by a discrete Hamiltonian:
\begin{gather}	
	-i\partial_z A = HA \,.
\end{gather}
Here, $A=(a_1,\ldots,a_n)^T$ is the state vector comprised of the individual
waveguide amplitudes $a_n$, and the symmetric square matrix
$H=(\delta_{m-1,n}+\delta_{m+1,n})C_n+\delta_{m,n}\beta_n$  contains the
individual propagation constants $\beta_n$ of the lattice sites on its main
diagonal, and the nearest neighbor coupling coefficients
$C_n\equiv C_{n,n+1}=C_{n+1,n}$ in the secondary diagonals
\cite{DiscretizingNature}.

In contrast to the continuous regime \cite{SUSY-PRL}, the discrete Hamiltonian
of a photonic lattice can be factorized, in a systematic and efficient manner, by a number of
exact algebraic techniques \cite{MathBook}. Here we employ the so-called
QR-method to find an orthogonal matrix $Q$ and an upper triangular matrix $R$,
such that $H=H^{(1)}-\lambda_k=QR$. In subtracting its corresponding eigenvalue
$\lambda_k$, any state can thus be eliminated from the partner Hamiltonian
$H^{(2)}=RQ+\lambda_k$. Note that the ability to address arbitrary states,
regardless of their position within the spectrum, sets the QR factorization
apart from the Cholesky method \cite{MathBook}, which is reminiscent of the
continuous-index approach, and is only valid for positive-definite matrices,
i.e. the removal of the fundamental mode.

Having established discrete optical supersymmetry, the question naturally arises
as to how scattering phenomena can be described in such guided-wave
arrangements, without violating the underlying coupled-mode theory. To do so, we
here consider only scattering processes within the transmission bands of the
lattice structures as established through the tight-binding model. Along these
lines, this nearest-neighbor coupling can lead to transverse transport,
diffraction and even reflection/transmission of waves at inhomogeneities in
extended lattices
\cite{AnomalousDiffraction,FresnelsLaws,SukhorukovReflectionless,SzameitReflectionless}.
In particular, owing to the periodic relation between transverse momentum and
propagation constant (in the propagation band), wave packets occupying the
vicinity of the diffraction relation's inflection points are known to propagate
without significant broadening \cite{AnomalousDiffraction}. In our experiments,
we will use this particular regime to probe the scattering properties of SUSY
photonic lattices.

\begin{figure}[h]
	\centering
	\includegraphics[width=\linewidth]{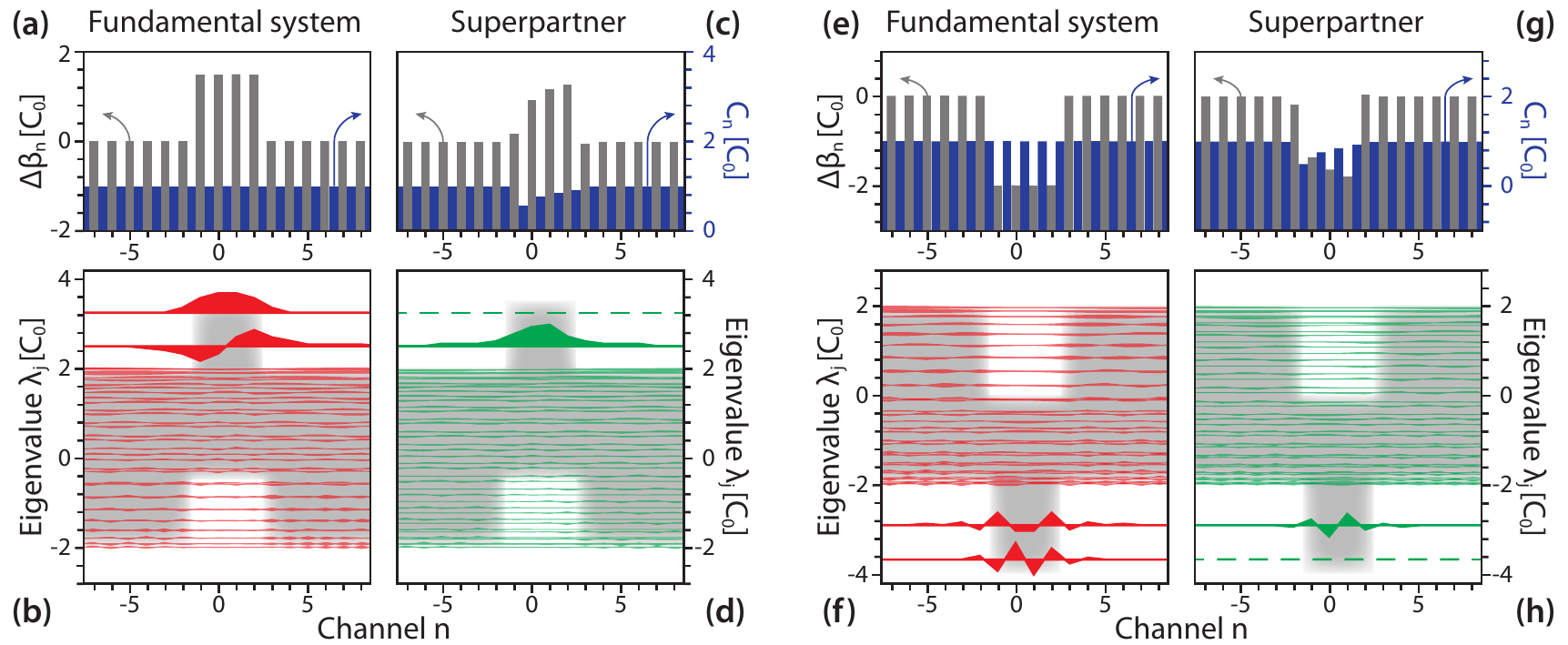}
	\caption{(Color online) (a) Positively detuned defect
	domain ($\Delta\beta=1.5\,C_0$) embedded within a homogeneous
	background lattice. Shown are the propagation constants $\beta_n$ of the individual
	lattice sites (gray bars, left scale), and the coupling coefficients
	$C_n$ between them (blue bars, right scale).
	(b) Lattice modes for the case of 18 waveguides on either side of the defect,
	i.e. 40 lattice sites in total. The mode profiles of the defect and radiation
	states are depicted as a function of the waveguide numbers, while the vertical
	position of each mode indicates the respective eigenvalue $\lambda_j$. The
	defect domain supports two bound states residing above the propagation band.
	(c) Superpartner lattice obtained by removing the fundamental defect state via
	QR factorization, and (d) its single bound mode residing above the band.
	(e) Negatively detuned defect domain ($\Delta\beta=-2\,C_0$) embedded within
	a homogeneous background lattice, and (f) its lattice modes. The defect domain
	supports two bound states residing below the propagation band.
	(g) Superpartner lattice obtained by removing the lowest defect state via
	QR factorization, and (h) its single bound mode residing below the band.}
	\label{fig1PositiveNegative}
\end{figure}

Note that, strictly speaking, supersymmetry is a global system property.
Appending any number of waveguides to either side of a finite array invariably
distorts the eigenvalue spectrum. As a consequence, one cannot simply immerse
two SUSY partner structures in identical homogeneous background lattices to
provide an interface for scattering experiments. Instead, the construction has
to be based on the entire fundamental system, including the uniform side
regions. The central domain then constitutes a defect, capable of supporting
localized modes. Figure \ref{fig1PositiveNegative}(a) shows such a defect domain
comprised of four identical waveguides, when their propagation constants have
been elevated by $1.5\times$ the coupling coefficient $C_0$ of the background
lattice: $\Delta\beta=1.5\,C_0$. The corresponding modes for an overall system
stretching 18 waveguides on either side of the defect, i.e. 40 lattice sites in
total, are shown in Fig. \ref{fig1PositiveNegative}(b). Most of their respective
eigenvalues $\lambda_j$ fall within the propagation band $\pm2\,C_0$. The only
exceptions are two localized states residing in the semiinfinite gap above the
band. SUSY transformations can now be employed to eliminate specific bound
states. The resulting partner system is largely similar to the fundamental one,
as deviations from the background lattice are confined to the area formerly
occupied by the now absent mode. Figure \ref{fig1PositiveNegative}(c) displays the
structure of the partner lattice obtained by QR factorization. Note that the
SUSY transformation yields joint modifications in both the propagation constants
and coupling coefficients, and leads to an asymmetric lattice. Yet,
supersymmetry does not introduce a sense of directionality: Reversing the
direction of the transverse coordinate would yield an equally valid
superpartner. Figure \ref{fig1PositiveNegative}(d) shows the corresponding lattice modes,
including the single localized defect state. While the density of states in the
propagation band depends on the extent of the surrounding lattice, the actual
wave dynamics in the vicinity of the defect domain remain unaffected by such
boundary effects. In that sense, the background lattice provides a continuum of
propagating waves, and can be truncated (or expanded) at will, as was done here
to maintain the overall system size of 40 waveguides. Similarly, the QR
factorization algorithm allows for the removal of localized states residing on negatively
detuned defect domains, even though their eigenvalues lie within the Bragg gap
of the surrounding lattice (see Fig. \ref{fig1PositiveNegative}(e-h)).

\begin{figure}
	\centering
	\includegraphics[width=.6\linewidth]{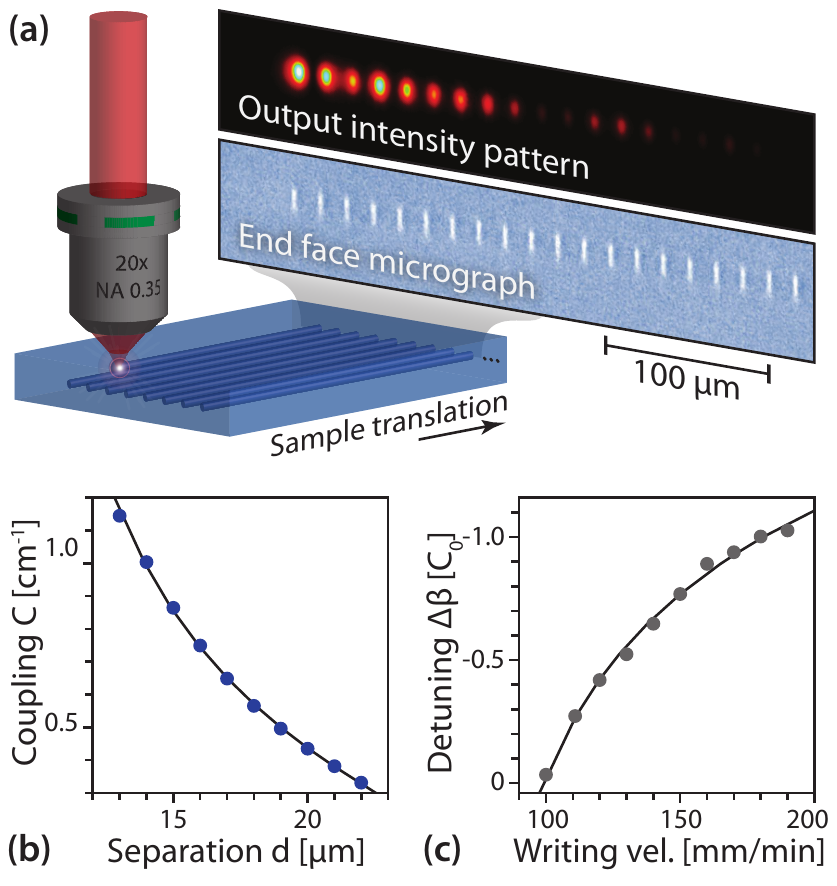}
	\caption{(Color online) (a) Schematic of the waveguide inscription method. The
		inserts show a micrograph of the sample end face, and a typical output
		intensity pattern observed at $633\eh{nm}$. (b) Dependence of the coupling
		coefficient $C$ on the waveguide separation $d$. (c) Influence of
		the writing velocity on the detuning $\Delta\beta$ of a
		waveguide with respect to a reference guide inscribed at $100\eh{mm/min}$.}
	\label{figWriting}
\end{figure}

\begin{figure}
	\centering
	\includegraphics[width=.6\linewidth]{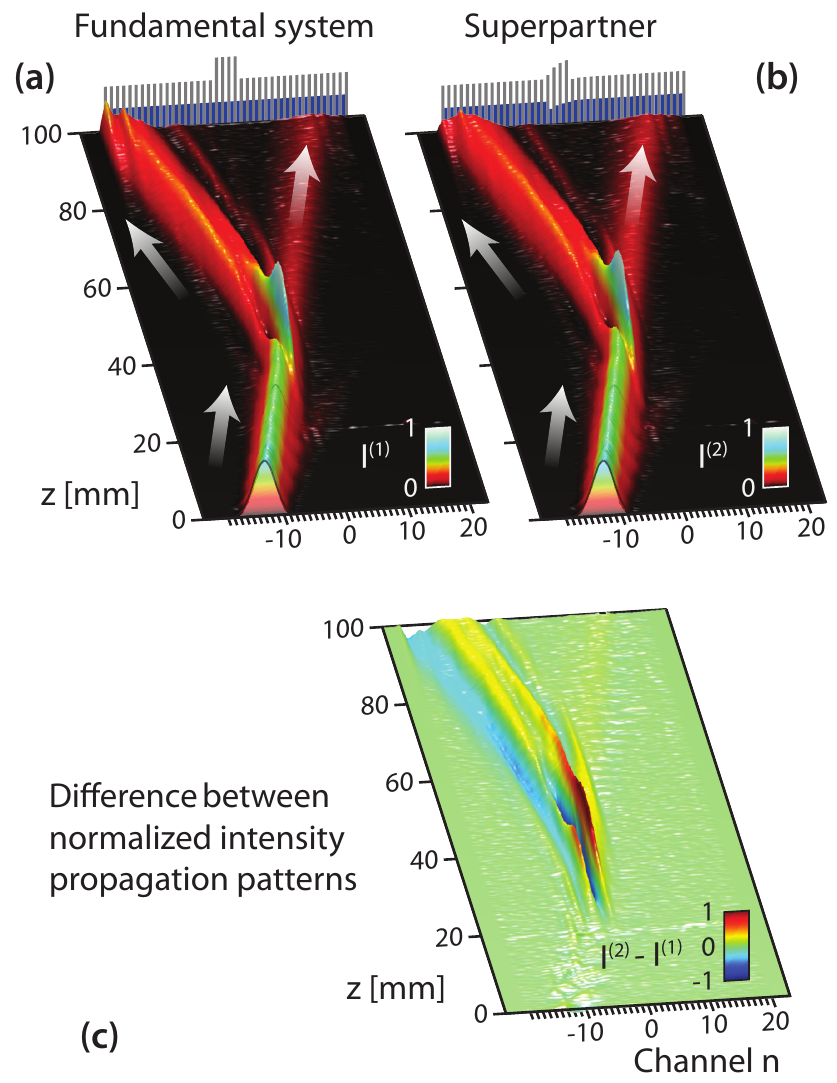}
	\caption{(Color online) (a) Evolution of a confined wave packet 
		upon partial reflection at a positively detuned defect domain
		($\Delta\beta=\,C_0$). Plotted is the intensity in the individual
		lattice sites $n$ along the longitudinal coordinate $z$. (b) Corresponding
		evolution in the superpartner lattice. (c) Observed quantitative difference
		between the patterns arising from a differential Goos-H\"anchen-like shift.
		The patterns in (a,b) are normalized to the same input beam intensity, and the
		values associated with the color bars are
		compatible in all subfigures.}
	\label{figScatteringExperiment}
\end{figure}

To experimentally study the scattering properties of discrete supersymmetric
arrangements, we employed the femtosecond laser direct writing technique
\cite{WaveguideReview} (see Fig. \ref{figWriting}(a)). By choosing appropriate
writing velocities and waveguide spacings, photonic lattices with the desired
propagation constants and coupling coefficients (see Fig. \ref{figWriting}(b,c))
were inscribed in $10\eh{cm}$ long fused silica samples. A confined,
diffraction-free probe beam was generated by placing the weakly focused
excitation beam (wavelength $633\eh{nm}$) in the background lattice, and
slightly tilting the sample by an angle corresponding to the inflection point of
the lattice band \cite{AnomalousDiffraction}. By virtue of the fluorescent
properties of our waveguides \cite{QuasiIncoherent}, we directly observed the
evolution of the beam and its interaction dynamics with the defect regions. In
order to allow for a quantitative comparison between the observed propagation
patterns (irrespective of the different waveguide positions in the superpartner
lattices), we numerically extracted the intensity distributions $I_n(z)$ of the
individual channels. Figure \ref{figScatteringExperiment} uses this type of
intensity plot to highlight the peculiar scattering behavior of SUSY photonic
lattices. Shown are the propagation patterns observed in the case of a
positively detuned defect domain with $\Delta\beta=\,C_0$ (Intensity $I^{(1)}$,
see Fig. \ref{figScatteringExperiment}(a)) and its superpartner (Intensity
$I^{(2)}$, see Fig. \ref{figScatteringExperiment}(b)). In addition to the
obvious visual similarity of these two scattering processes, a closer
examination reveals a difference between these two intensities
$I^{(2)}-I^{(1)}$, a direct outcome of the phase differences involved. Since the
incoming wave packet encounters a less abrupt change of lattice parameters in
the supersymmetrically deformed defect region, it can penetrate slightly
further, resulting in a longer Goos-H\"anchen-like displacement along $z$.
Nevertheless, both have the same asymptotic shape and carry the same fraction of
the overall intensity, as dictated by SUSY. Our experiments clearly demonstrate
that the intensity reflection/transmission coefficients of these two partner
structures are indeed identical.

\begin{figure}
	\centering
	\includegraphics[width=.6\linewidth]{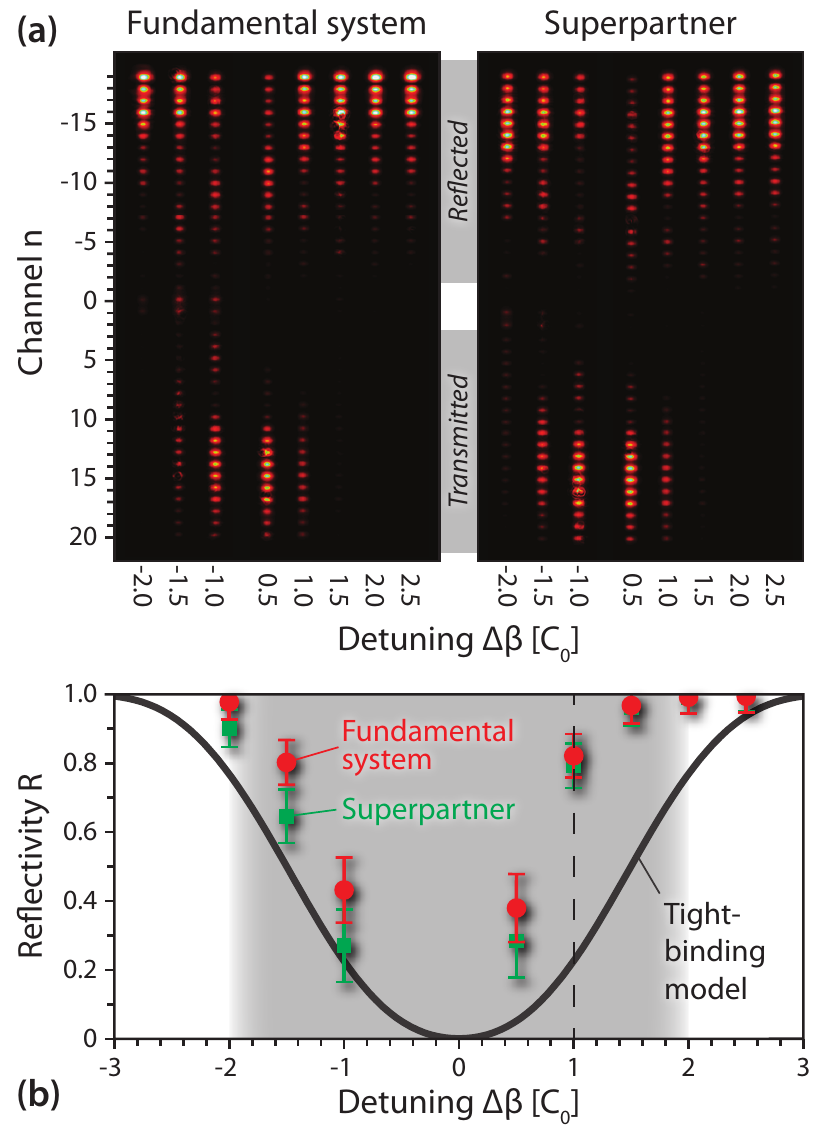}
	\caption{(Color online) Reflection/transmission from defect domains
	and their respective superpartners for different values of detuning. 
	Observed output intensity distributions at $z=100\eh{mm}$ in the 
	fundamental lattices (left) and SUSY partners (right). The reflected and
	transmitted parts of the probe beam are located at $n\leq-2$ and $n\geq+3$, respectively.
	(b) Values of the reflectivity $R$ extracted from the output patterns. For
	comparison, the theoretical band structure is shown as a solid graph. The
	dashed vertical line indicates the data points corresponding to the propagation
	patterns shown in Figs. \ref{figScatteringExperiment}(a).}
	\label{figScatteringGraph} 
\end{figure}

In a second set of experiments, we realized a number of different defect
arrangements with positive ($\Delta\beta=0.5,1.5,2.0,2.5\,C_0$) as well as
negative detunings ($\Delta\beta=-2.0,-1.5,-1.0\,C_0$). Their superpartners were
synthesized by removing the respective most strongly localized defect states,
i.e. the modes with the highest (lowest) eigenvalue for the positive (negative)
defects. We then observed the output intensity distributions at the sample end
faces. Compared to the fluorescence method, this measurement is subject to a
significantly lower background, and the modal intensities can be extracted with
high fidelity to obtain quantitative information about the
reflection/transmission behavior. The left side of Fig.
\ref{figScatteringGraph}(a) shows the end face images, normalized with respect to their individual overall
intensities. The reflected parts of the input beam occupy the lattice sites with
$n\leq-2$, whereas the waveguides $n>2$ carry the transmitted fraction. For the
lattice size in our experiments, the reflectivity is therefore given by
$R=\sum_{n=-19}^{-2}I_n/I_\mathrm{total} $. The values of $R$ obtained from the
measurements for the defects and their respective superpartners closely match
one another within their margins of error. As predicted by the tight binding model (see right side of Fig.
\ref{figScatteringGraph}(a)), the reflectivity increases with the
relative strength of the defect. This can be easily understood by considering that a
sufficiently detuned region can effectively separate the surrounding lattice: If
it does not support any states within the propagation band, the defect ideally
poses an impenetrable barrier to light contained in the scattered states. As the
detuning decreases, so does the fraction of domain modes that lie outside the
band. Naturally, at zero detuning, the domain would become indistinguishable
from the lattice, and exhibit vanishing reflectivity. Deviations from the
theoretical prediction illustrate the fact that the experimental system departs
from the ideal tight-binding configurtation, e.g. by displaying higher-order
couplings \cite{HigherOrderCoupling}, which subtly change the shape of the
diffraction relation as well as it's associated density of states. Nevertheless, lattices related by
SUSY-transformations clearly resemble one another with respect to their
scattering properties. Along these lines, we note that SUSY notions also provide
a new, and possibly more intuitive, perspective to previous studies on
reflectionless potentials \cite{SukhorukovReflectionless,SzameitReflectionless}.

In conclusion, we implemented discrete supersymmetric optical structures in
femtosecond laser-written waveguide photonic lattices, and experimentally
studied the scattering characteristics of such systems. Our findings constitute
the first observation of SUSY scattering behavior in either continuous or
discrete systems, and illustrate how lattice-type arrangements readily allow for
the implementation and observation of optical supersymmetry. 
Of interest will also be an extension of these experiments to non-Hermitian
SUSY-synthesized systems, which have recently been predicted to offer
particularly unusual scattering behavior \cite{LonghiX1,LonghiX2}.

\section*{Acknowledgements}
The authors gratefully acknowledge financial support from NSF (grant
ECCS-1128520), AFOSR (grants FA9550-12-1-0148 and FA9550-14-1-0037), the German Ministry of Education and
Research (Center for Innovation Competence program, grant 03Z1HN31), Thuringian
Ministry for Education, Science and Culture (Research group Spacetime, grant no.
11027-514) and the German-Israeli Foundation for Scientific Research and
Development (grant 1157-127.14/2011). MH was supported by the German National
Academy of Sciences Leopoldina (grant LPDS 2012-01).

\pagebreak
\section*{Figures}

\begin{thebibliography}{99}

\bibitem{RamondSUSY}
		P. Ramond,
		``Dual Theory for Free Fermions,''
		Phys. Rev. D \textbf{3}, 2415–2418 (1971).
	
	\bibitem{CooperKhare}
		F. Cooper, A. Khare, U. Sukhatme, and A. Khare,
		``Supersymmetry and quantum mechanics,''
		Phys. Rep., \textbf{251}, 267–385 (1995).
		
	\bibitem{SUSYQuantumWell}
		J. Radovanovi\'{c}, V. Milanovi\'{c}, Z. Ikoni\'{c}, and D. Indjin, 
		``Quantum-well shape optimization for intersubband-related electro-optic modulation properties,''
		Phys. Rev. B \textbf{59}, 5637 (1999).
	
	\bibitem{BaiCitrin}
		J. Bai, and D. S. Citrin, 
		``Supersymmetric optimization of second-harmonic generation in mid-infrared quantum cascade lasers,''
		Opt. Express \textbf{14}, 4043–4048 (2006).
	
	\bibitem{SUSY-PRL}
		M.-A. Miri, M. Heinrich, R. El-Ganainy, and D. N. Christodoulides,
		``Supersymmetric Optical Structures,''
		Phys. Rev. Lett. \textbf{110}, 233902 (2013).
		
	\bibitem{SUSY-PRA}
		M.-A. Miri, M. Heinrich, and D. N. Christodoulides,
		``Supersymmetry-generated complex optical potentials with real spectra,''
		Phys. Rev. A \textbf{87}, 043819 (2013).
		
	\bibitem{UkraineSUSY}
		H. P. Laba and V. M. Tkachuk,
		``Quantum-mechanical analogy and supersymmetry of electromagnetic wave modes in planar waveguides,''
		Phys. Rev. A \textbf{89}, 033826 (2014) 
		
	\bibitem{SUSY-MC}
		M. Heinrich, S. St\"{u}tzer, M.-A. Miri, R. El-Ganainly, S. Nolte, A. Szameit, and D. N. Christodoulides,
		``Supersymmetric mode converters,''
		Nature Commun. \textbf{5}, 3698 (2014).
	
	\bibitem{LonghiReflection}
		S. Longhi and G. Della Valle,
		``Transparency at the interface between two isospectral crystals,''
		Europhys. Lett., \textbf{102}, 40008 (2013).
	
	\bibitem{LonghiInvisible1}
		S. Longhi,
		``Invisibility in non-Hermitian tight-binding lattices,''
		Phys. Rev. A, \textbf{82}, 032111 (2010).
		
	\bibitem{LonghiInvisible2}
		S. Longhi and G. Della Valle,
		``Invisible defects in complex crystals,''
		Ann. Phys. \textbf{334}, 35–46 (2013).
		
	\bibitem{MidyaInvisible}
		B. Midya,
		``Supersymmetry-generated one-way-invisible PT-symmetric optical crystals''
		Phys. Rev. A \textbf{89}, 032116 (2014).
		
	\bibitem{SUSY-TO}
		M.-A. Miri, M. Heinrich, and D. N. Christodoulides,
		``Supersymmetric transformation optics,''
		Optica \textbf{1}, 89 (2014).
		
	\bibitem{WaveguideReview}
		A. Szameit and S. Nolte,
		``Discrete optics in femtosecond-laser-written photonic structures,''
		J. Phys. B \textbf{43}, 163001 (2010).
	
	\bibitem{QuasiIncoherent}
		A. Szameit, F. Dreisow, H. Hartung, S. Nolte, A. Tünnermann, and F. Lederer,
		``Quasi-incoherent propagation in waveguide arrays,''
		Appl. Phys. Lett. \textbf{90}, 241113 (2007).
		
	\bibitem{DiscretizingNature}
		D. N. Christodoulides, F. Lederer, and Y. Silberberg,
		``Discretizing light behaviour in linear and nonlinear waveguide lattices.,''
		Nature \textbf{424}, 817–23 (2003).
		
	\bibitem{MathBook}
		L. Hogben,
		\textit{Handbook of Linear Algebra},
		(Chapman \& Hall/CRC, 2006).
		
	\bibitem{AnomalousDiffraction}
		T. Pertsch, T. Zentgraf, U. Peschel, A. Br\"auer, and F. Lederer,
		``Anomalous Refraction and Diffraction in Discrete Optical Systems,''
		Phys. Rev. Lett., \textbf{88} 093901 (2002).
		
	\bibitem{FresnelsLaws}
		A. Szameit, H. Trompeter, M. Heinrich, F. Dreisow, U. Peschel, T. Pertsch, S.
		Nolte, F. Lederer, and A. T\"{u}nnermann, ``Fresnel’s laws in discrete optical
		media,'' New J. Phys., \textbf{10}, 103020 (2008).
			
	\bibitem{SukhorukovReflectionless}
		A. A Sukhorukov,
		``Reflectionless potentials and cavities in waveguide arrays and coupled-resonator structures,''
		Opt. Lett., \textbf{35}, 989–991 (2010).
	
	\bibitem{SzameitReflectionless}
		A. Szameit, F. Dreisow, M. Heinrich, S. Nolte, and A. A. Sukhorukov,
		``Realization of Reflectionless Potentials in Photonic Lattices,''
		Phys. Rev. Lett., \textbf{106}, 193903 (2011).
		
	\bibitem{HigherOrderCoupling}
		F. Dreisow, A. Szameit, M. Heinrich, T. Pertsch, S. Nolte and A. T\"{u}nnermann,
		``Second-order coupling in femtosecond-laser-written waveguide arrays,''
		Opt. Lett. \textbf{33}, 2689-2691 (2008).
		
	\bibitem{LonghiX1}
		S. Longhi and G. Della Valle,
		``Optical lattices with exceptional points in the continuum,''
		Phys. Rev. A \textbf{89}, 052132 (2014).
		
	\bibitem{LonghiX2}
		S. Longhi,
		``Bound states in the continuum in PT-symmetric optical lattices,''
		Opt. Lett. \textbf{39}, 1697 (2014).	
\end{thebibliography}
\end{document}